# Observing stellar bow shocks

A.C. Sparavigna[a] and R. Marazzato[b]

[a] *Department of Physics, Politecnico di Torino, Torino, Italy*
[b] *Department of Control and Computer Engineering, Politecnico di Torino, Torino, Italy*

**Abstract**
For stars, the bow shock is typically the boundary between their stellar wind and the interstellar medium. Named for the wave made by a ship as it moves through water, the bow shock wave can be created in the space when two streams of gas collide. The space is actually filled with the interstellar medium consisting of tenuous gas and dust. Stars are emitting a flow called stellar wind. Stellar wind eventually bumps into the interstellar medium, creating an interface where the physical conditions such as density and pressure change dramatically, possibly giving rise to a shock wave. Here we discuss some literature on stellar bow shocks and show observations of some of them, enhanced by image processing techniques, in particular by the recently proposed AstroFracTool software.

**Keywords**: Shock Waves, Astronomy, Image Processing

**1. Introduction**
Due to the presence of space telescopes and large surface telescopes, equipped with adaptive optics devices, a continuously increasing amount of high quality images is published on the web and can be freely seen by scientists as well as by astrophiles. Among the many latest successes of telescope investigations, let us just remember the discovery with near-infrared imaging of many extra-solar planets, such as those named HR 8799b,c,d [1-3]. However, the detection of faint structures remains a challenge for large instruments. After high resolution images have been obtained, further image processing are often necessary, to remove for instance the point-spread function of the instrument. This is an essential processing, which helped determining the presence of planets revolving about stars. Besides the well-known enhancement techniques, such as the Richardson and Lucy deconvolution, [4,5], many other processing methods are involved in astrophysical researches. Among the others, those based on the MCS deconvolution algorithm revealed spectacular gravitational lensing effects, such as that at cluster Abell 2218, where distant blue galaxies are squished into a circular thin ring around the middle of the cluster [6-10].
Other interesting objects are the nursery of stars, dense nebulae, billowing clouds of gas and dust. The filamentary structures of these clouds can be enhanced by image processing, to appreciate the detailed directions of the streaming flows. A beautiful example of cometary structures can be observed in NGC5189 (see Fig.1). In the nurseries, young stars are emitting strong solar winds. In the space near these stars, the blowing stellar wind can create regions where the conditions of density and pressure change dramatically, giving rise to shock waves. These shock waves are also known as bow shocks, as the waves made by a ship as it moves through water.
The literature on the stellar bow-shocks is now rapidly increasing, due to many research projects based on the use of infrared observations, able to reveal the emission of energy from the bow shocks, distinguished from the stars themselves. Let us consider for instance the MIRIAD project, based on the use of Spitzer Space telescope: this project has as primary aim to probe the material distribution in the extended circumstellar envelopes of evolved stars. Ref.[11] shows a bow shock at the interface of the interstellar medium and the wind of the moving star R Hydrae. According to [11], this discovery exemplifies the potential of Spitzer as a tool to examine the detailed structure of extended far-IR nebulae around bright central sources.

As we shall see, besides Spitzer, other space telescopes, such as Akari, and surface IR telescopes, are able to reveal bow shocks created in the interstellar medium. Here, we discuss some observations of stellar bow shocks, chosen among the scientific literature. In particular, we discuss two observations that we, and the readers, can easily repeat using images freely published on the web. In the enhanced images, obtained with our recently proposed processing technique, we have a better view of the bow shocks, because the software is able to enhance many details in images, such as edges and faint objects (an example is shown in Fig.1) [12,13]. Before reporting the bow shocks observations, let us shortly discuss their physics.

## 2. Shock waves

The base theory of shock waves can be considered a part of fluid mechanics modelling. We find shock waves in bullet motions, explosions, as well as in astrophysics. To have a shock wave, we need a motion comparable or exceeding the sound speed. Shocks start to occur in the limit where the pressure variation cannot be considered as small. As a simple example of how these waves arise, consider the one-dimensional flow of gas in a small-diameter cylindrical tube, which is fitted with a piston at one end and closed at the other end. As the piston is moved into the fluid, the fluid starts to compress. Information about this rise in pressure propagates away from the piston at the sound speed $c_s$ of the fluid. If the piston speed $v_p$ is greater than the sound speed, the pressure continues to build in front of the piston with the gradient in pressure becoming steeper and steeper. The edge of the pressure hump (the shock) moves down the tube at speed $v_s$. We can define the Mach number as $M=v_s/c_s$. Since sonic disturbances travel in the gas at the local sound speed, and since the gas immediately ahead of the piston is moving with the piston speed, the sound waves emanating from each point along the piston path travel at a speed of $v_p + c_s$. The waves converge to form a shock front, travelling along the tube at supersonic speeds.

Landau and Lifshitz discussed the shock waves in their book on fluid mechanics [14]. Their discussion of shock uses a frame of reference where it is at rest. At the stationary shock, a set of equations, the Rankine- Hugenoit equations, relates the density, pressure and flux conditions on either side of it. These equations are characterized by the conservation of three quantities: the mass, the flux of momentum and the specific total energy (for more details, see [15]). These conditions can be discussed to obtain the relative conditions before and after the shock. As in Ref.15, we can distinguish isothermal from adiabatic shocks: in fact, real shocks have conditions between adiabatic and isothermal shocks.

## 3. Shocks in the space

The equations of hydrodynamics are valid if the particle mean-free path is small in comparison with characteristic lengths involved in the problem. In the space, where the medium is so rarefied, that conditions for shock seem to be impossible. Nevertheless shock waves exist. In fact, the interstellar medium and the solar wind contain plasma. For a plasma in a magnetic field, the hydrodynamic approximation can be used even in those cases when this criterion is not fulfilled, because, in a magnetic field there is a second characteristic length, the Larmor radius [16]. If this length is much less than the characteristic size of the system, the equations of magneto-hydrodynamics (MHD) will be valid. The theory is of fluid can be then changed in the magneto-hydrodynamic fluid theory, to have the so-called Alfven Magnetosonic waves. MHD shocks are similar to adiabatic shocks, but with the addition of a magnetic field term and one more conserved quantity, the magnetic flux.

Numerous examples of cosmic shocks exist. The earth and other planets have magnetic fields. The solar wind, mostly composed by protons and electrons, encounters an MHD shock at the earth's Bow Shock, after which particles move in the earth's magnetic field [15]. Another shock is caused by the solar wind, at the so-called heliopause, where the sun's particles flow out into the interstellar medium. Larger stars, having dense winds, produce great shocks. An example of a cosmic shock is a Herbig-Haro object [15]. It is observed as small patches of nebulosity associated with a newly born star, formed when the gas ejected by the star collides with clouds of gas and dust. Herbig–Haro objects are ubiquitous in star-forming regions that can evolve visibly over short timescales as they move rapidly away from their parent star into the gas clouds in interstellar space (see for instance, [17]). Spectacular shock waves are also visible in the form of supernova remnants.

**4. Stellar bow shock**
Stellar wind bow shocks are structures due to the supersonic passage of wind-blowing stars [18]. They condense the interstellar matter into thin shells, which may be revealed by their post-shock emission or by scattered light. By means of these structures, stellar winds, that might otherwise go undetected, can be studied. A theory, that of momentum-supported bow shocks, is discussed in Ref.18: this theory was first developed by Baranov, Krasnobaev and Kulikovskii [19], who were motivated by solving the problem of solar wind and local interstellar medium interaction. They considered the collision of an isotropic stellar wind with a uniform ambient medium, including the supersonic motion of the star with respect to that medium, and solved numerically the equation to have the shape of the bow shock. In [18], the author presents a formulation of the problem, that allows to obtain simple, exact solutions for all quantities in the numerical model of BKK, such as the shell's shape and the mass and velocity distributions within it.

Where the stellar wind and the ambient medium collide head-on, it is possible to define the radius of the starting point of the shell, found by balancing the ram pressure of the wind and the ambient medium [18]. This radius is given by $R_0 = (\dot{m}_w V_w / 4\pi \rho_a V_S^2)^{1/2}$, where $\dot{m}_w, V_w$ is the mass-loss rate and the constant speed of the stellar wind, $V_S$ the speed of the star and $\rho_a$ the density of the uniform ambient medium. This standoff distance sets the length scale of the shell. The shape is a universal function. Because the bow shock depends upon both the stellar wind's and ambient medium's properties, stellar wind bow shocks may be very useful probes of both. According to [18], it is then possible to recover the mass and momentum flux functions and thus the stellar mass-loss. The local ambient density can be derived from equations too.

In the following, some examples of observed stellar bow shocks.

**4.1 Betelgeuse**
Akari Space Telescope observations of Betelgeuse, the bright red supergiant star located in the constellation Orion, 640 light-years from Earth, show the star making a bow shock as it crosses the interstellar medium. At [20,21], three-colour composite images of Betelgeuse and its surroundings, taken at 65, 90 and 140 micrometers, are proposed: Betelgeuse travels through the interstellar medium, creating a bow shock. As previously discussed, it is not the star itself that creates the bow shock, but rather the interaction of its stellar wind with the gas in the interstellar medium. This wind warms up the surrounding gas releasing light in the infrared. The shape of a bow shock can be calculated exactly using a shock theory based on the conditions of the shock [18]. By analyzing the shape of the bow shock around Betelgeuse, researchers figured out that there is a strong flow of the interstellar medium around Betelgeuse: the interstellar medium, originated from star forming regions in the Orion's Belt is flowing at 11 km/s and Betelgeuse is crossing this flow at 30 km/s, while spewing out its wind at 17 km/s. The combined motion of Betelgeuse and its wind is behaving as a ship crossing a river [21].

**4.2 In the Orion Nebula again**
At [22], an image shows further evidence of bow shock existence from dense gases and plasma in the Orion Nebula. In this case, it is the Hubble Space Telescope revealing the structures residing within the intense star-forming region which is the Great Nebula in Orion. One such structure is the bow shock around the star, LL Orion. Again, this star emits a solar wind. The material in the fast wind from LL Orion collides with slow-moving gas evaporating away from the center of the Orion Nebula. The surface where the two winds collide has a clearly crescent bow shape. The presence of a cometary tail after the star creates an image looking like a bow with its arrow.

Let us note that, unlike the bow-shock made by a ship, an interstellar bow shock is a three-dimensional structure. The structure has a clear boundary on the side facing away from LL Orion and it is diffuse on the side closest to the star: this is a characteristic common to many bow shocks [22]. A second bow shock can be seen around a star near LL Orion. The image in Ref.22, is composite and obtained with specific filters to represent oxygen, nitrogen, and hydrogen emissions.

**4.3 RCW 49 nebula**
One of the most prolific birthing place in our galaxy, the nebula RCW 49, was observed in high detail for the first time by the Spitzer Space Telescope. Located 13,700 light-years away in the Centaurus constellation, this is a dusty stellar nursery with more than 2,200 stars. Many of the stars cannot be seen at visible wavelengths but viewed with Spitzer's infrared camera. Ref.23 describes the structure of the nebula. The interstellar medium structures are dominated by two large cavities. The first, blown out to the west, contains the massive young cluster Westerlund 2, and the second is an enclosed bubble around the Wolf-Rayet star WR 20b. In Ref.[23] the researches show three bow shocks associated with RCW 49.
These bow-shocks are interesting: none of them point directly back toward the central cluster. This could be a consequence that the expanding bubbles driven by Westerlund 2 and the Wolf-Rayet stars are interacting and then giving a non-radial components to the flows. It is also possible that the bow shock driving stars have large orbital motions relative to the dynamic interstellar medium.
In Fig.2, we show a detail of an original high-quality image, obtained from a web page of the University of Wisconsin-Madison [24], containing two of the bow shocks described in Ref.23. On the right of the same Figure 2, it is shown the image obtained enhancing the edges of interstellar clouds by means of AstrofracTool. Bow shocks form Ref.23 are marked with square boxes. The enhanced image reveals another bow structure (encircled) which is not considered in Ref.23. In our opinion, an investigation on images with higher resolution could be rather interesting to understand whether a bow shock is present or not.

**4.4 The Galactic Center**
We can observe bow shocks at the center of our Galaxy too.
Because of the high dust content along the line of sight to the Galactic Center, it is impossible to observe this region at ultraviolet and optical wavelengths, but the infrared array detectors combined with adaptive optics give astronomers a new way to explore this region. In the past years, many investigation with high-resolution near-infrared imaging have been carryed out of the central few light years of the Milky Way, in particular in the vicinity of the compact radio source SgrA*. This is the most likely counterpart of the supposed black hole, occupying the center of the Galaxy. The accretion of gas onto the black hole would release energy to power the radio source [25,26].
The Galactic Center also contains a number of young clusters of recently formed hot and massive stars. Many stars are old red main sequence stars. The existence of relatively young stars raise problems because, it was expected that the tidal forces from the central black-hole prevent their formation [25]. This paradox of youth is even more remarkable for stars that are on very tight orbits around Sagittarius A* [26]. Proposed explanation are that stars were formed in a massive star cluster offset from the Galactic Center, that eventually migrated to its current location once formed, or that stars formation happens within a rather compact accretion disk around the central black-hole. In these dense regions, scientists of the Gemini Observatory, analyzing images obtained with the Hawaii's adaptive optics on Gemini North, found that an object, IRS-8, known to be a strong infrared source for almost two decades, has an infrared largely in the form of a spectacular bow shock surrounding a central star [27]. It is the heated dust in the bow shock which is radiating, because it is compressed by the shock and because it is absorbing radiation from the hot neighbouring stars, including the IRS-8's central star itself. This is an interesting example, showing how new observations can discriminate the true nature of an infrared source.

Other observations of the Galactic Center have been obtained from the AO imager/spectrometer NAOS/CONICA, mounted on the Yepun Telescope, Chile. As claimed in Ref.28, this is the ideal instrument for tackling all the basic questions concerning the Galactic Center. In Ref.29, using this device, researchers obtained images of a bow shock in the region near IRS7. In Figure 3, we show this region, as seen in the image from the Gemini South Telescope [30]. Let us enhance the image with AstrofRacTool: the image on the right is obtained. Note that the structure of the Galactic Center is strongly enhanced. The bow shock reported in Ref.29 is now clearly visible, marked by a square box.

**Conclusions**

We devoted this paper to the discussion of bow shocks in the interstellar medium created by the wind of moving stars. We discussed some observations of stellar bow shocks, chosen among the scientific literature. The two examples proposed in 4.3 and 4.4 are important in our opinion, because the reader can easily repeat the investigation using images, which can be freely downloaded from the web. In the enhanced images, obtained with a recently proposed processing method, we have a better view of the bow shocks.

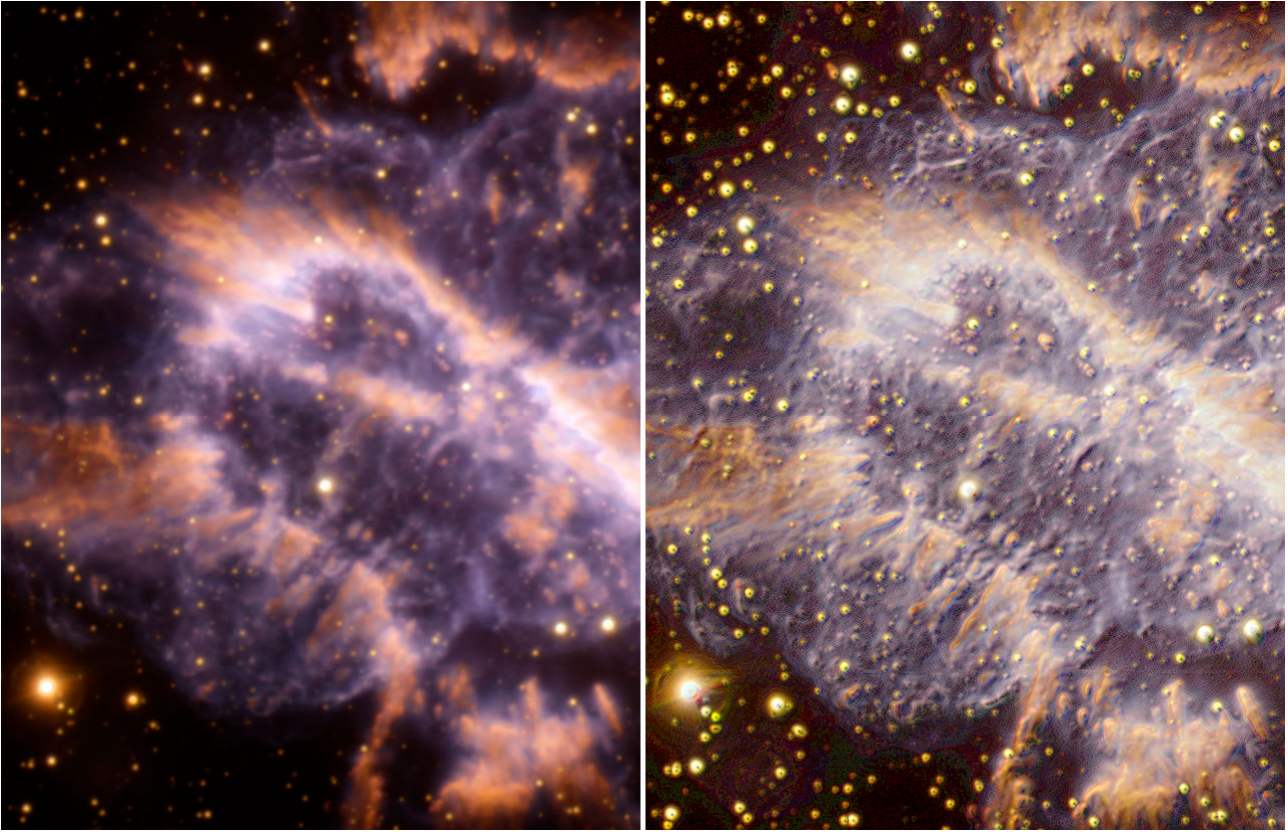

Fig.1 Cometary structures in Ngc5189: on the left the original image from GEMINI web site and on the right the image as obtained after AstroFracTool and GIMP application.

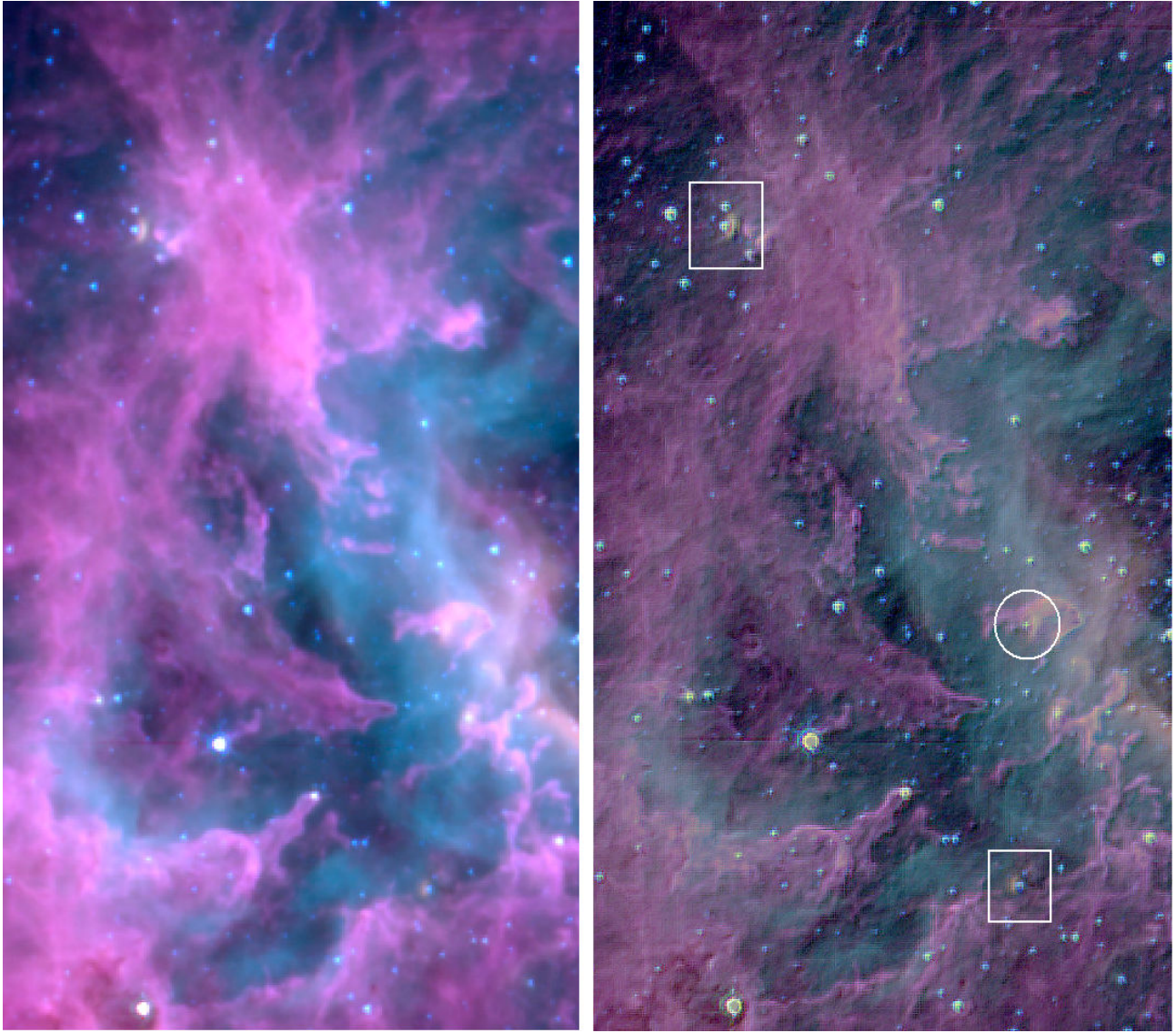

Fig.2 Bow shocks in RCW49: on the left the original image (credits: NASAJPL-Caltech, University of Wisconsin-Madison) and on the right, the same image enhance after AstroFracTool application. Bow shocks form Ref.23 are marked with square boxes. The enhanced image reveals a bow structure (encircled) which is not considered in Ref.23.

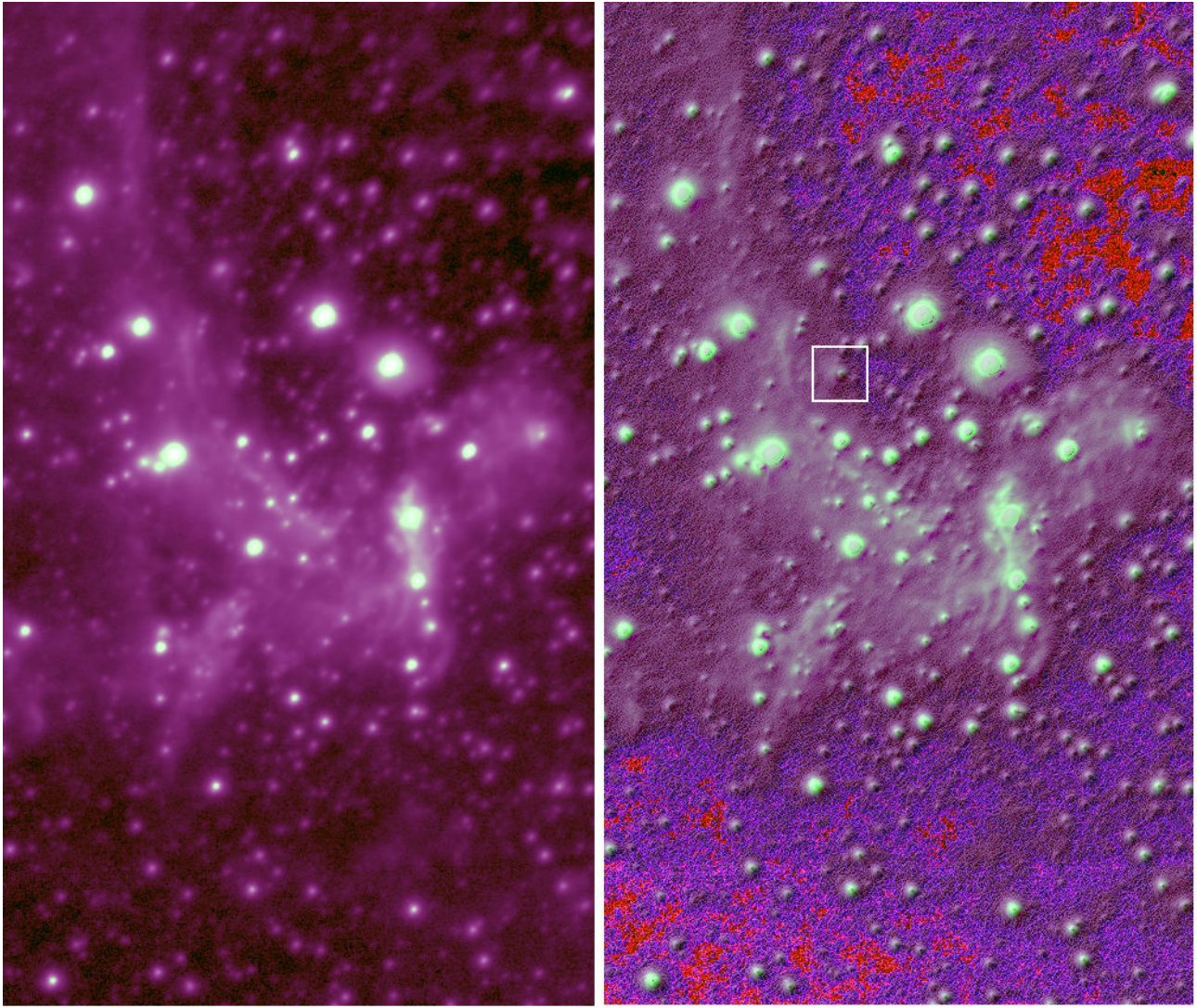

Fig,.3 The Galactic Center (credits: Gemini Observatory-AbuTeam-NOAO-AURA-NSF). This infrared image reveals the core of our Galaxy. On the right the same image after AstroFracTool processing. Note the enhancement of the structures. The bow shock is clearly visible, marked in the square box. This is the bow shock reported in Ref.29.